%% file: PG2130.tex
\documentclass[twocolumn]{aastex631}
\usepackage{array}
\usepackage{multirow}
\newcolumntype{P}[1]{>{\centering\arraybackslash}p{#1}}
\usepackage{graphicx}	
\usepackage{amsmath}	
\usepackage{amssymb}	
\usepackage{comment}
\usepackage{float}
\usepackage{newtxtext,newtxmath}
\usepackage{booktabs}
\usepackage[graphicx]{realboxes}
\usepackage{rotating}
\usepackage{nicefrac}
\newcommand{\angstrom}{\mbox{\normalfont\AA}}

\usepackage{amsmath} 

\received{November 21st, 2024}
\revised{March 17th, 2025}
\accepted{March 31st, 2025}

\shorttitle{Investigating the Variable Continuum Lags in PG~2130+099}
\shortauthors{Miller et al.}
\begin{document}

\title{Investigating the Variable Continuum Lags in PG~2130+099}
\author[0000-0001-8475-8027]{Jake A. Miller}
\affiliation{Wayne State University, Department of Physics \& Astronomy, 666 W Hancock St, Detroit, MI 48201, USA}
\affiliation{Texas A\&M University, Department of Physics \& Astronomy, 400 Bizzell St, College Station, TX 77845, USA}

\author[0000-0002-8294-9281]{Edward M. Cackett}
\affiliation{Wayne State University, Department of Physics \& Astronomy, 666 W Hancock St, Detroit, MI 48201, USA}
\author[0000-0002-2908-7360]{Mike Goad}
\affiliation{School of Physics and Astronomy, University of Leicester, University Road, Leicester, LE1 7RH, UK}
\author[0000-0003-0944-1008]{Kirk T. Korista}
\affiliation{Department of Physics, Western Michigan University, 1120 Everett Tower, Kalamazoo, MI 49008-5252, USA}


\begin{abstract}
Broadband photometric reverberation mapping (RM) provides a measure of the size of the continuum-emitting region in active galactic nuclei (AGN). Previous monitoring campaigns of PG~2130+099 disagree as to whether the continuum emitting region size is consistent with that predicted for a standard optically thick geometrically thin accretion disk. We present $\sim$6 months of observations from several robotic telescopes, providing the highest cadence and widest wavelength coverage photometric RM study of PG~2130+099 to date. Our results indicate that inferred size of the continuum-emitting region in PG~2130+099, like many recently observed AGN, {is larger than the} simplest predictions for an irradiated geometrically thin, optically thick accretion disk. We also perform a flux-flux analysis, finding a variable spectrum broadly consistent with a disk, and a constant component with enhanced $i$-band emission, potentially due to H$\alpha$. We find some evidence of increasing lag with luminosity, but previous lag measurements are too uncertain to be definitive.

\end{abstract}

\section{Introduction}
Active Galactic Nuclei (AGN) occur in the centers of galaxies when enough material surrounding a supermassive black hole (SMBH) is accreted. This infalling material forms an accretion disk which radiates like a blackbody due to magnetohydrodynamical friction, and is thought to be the central engine that powers other phenomena observed in the AGN. Thus, understanding how this material is accreted onto the SMBH is necessary for understanding how these objects evolve over time. 

Accretion disks are usually too compact and distant to be spatially-resolved, so we must use indirect observation methods. A prominent method for determining disk size is called continuum reverberation mapping (RM). This method assumes that a {compact} central ionizing source, usually assumed to be X-rays powered by Compton upscattering, irradiates the accretion disk \citep[e.g.,][]{collier1999, cackett2007}, which is thought to be geometrically thin and optically thick \citep{1973A&A....24..337S}. In this picture, the disk reprocesses this emission, absorbing the X-rays and re-emitting at longer wavelengths. At hotter regions closest to the {SMBH}, this will produce primarily ultraviolet radiation, while cooler regions further out will primarily produce optical emission. 

In this scenario, X-rays should drive all short-term variability, and thus the variability patterns of the ultraviolet and optical light curves should be smoothed and delayed versions of the X-ray light curve. Assuming the disk radiates like a multicolor blackbody, we can determine how the time lags should scale with wavelength. 

Continuum {RM} campaigns spanning a range of durations and sampling rates have produced results in general agreement to these predictions, although with some caveats. While the interband continuum delays have been observed to generally scale like $\lambda^{4/3}$, the implied disk sizes that result from the measured lags are roughly $\sim$3x bigger than simple analytical predictions \citep[e.g.,][]{shappee2014, edelson2015, fausnaugh2016, kara2021}. In addition, while the X-rays are thought to drive variability {at} longer wavelengths, they often exhibit a far weaker correlation than {is expected (\citealt[][]{breedt2009, edelson2019, santisteban2020}; although see \citealt{panagiotou2022})}. Finally, it is thought that the broad line region (BLR) contributes significantly to the observed optical emission, both from broad emission lines and {reprocessed continuum emission}. This last point has received significant attention in recent RM studies \citep{korista2001, korista2019, lawther2018, netzer2022,fausnaugh2016, cackett2018, guo2022, jiang2024}. 

Determining what fraction of the continuum lags are due to reprocessing from the disk, or from the BLR continuum is vitally important if we are to {correctly interpret} the implied accretion disk sizes. An object exhibiting unique BLR properties may shine some light onto this analysis. PG~2130+099 (also known as II~Zw~136 and Mrk~1513) is such an object, with {more than 20~years of monitoring data and a diverse range in campaign duration and sampling rates}. 

The first BLR {RM} study of PG~2130+099 {was by \citet{2000ApJ...533..631K}. They} found a {mean H$\alpha$ and H$\beta$} radius of $200^{+67}_{-18}$~light-days, with an SMBH mass of $14.4^{+5.1}_{-1.7}\times10^7M_\odot$. This is one of the largest BLR radii in their sample of 17 Palomar-Green quasars. {The sampling of the PG quasars was sub-optimal, with only 64~H$\beta$ observations over a span of 6~years leading to large uncertainties in the measured delays. They are also more luminous and vary on longer timescales, additionally obfuscating the true delay.} This result was reanalyzed by \citet{peterson2004_RMDB} and found to still be rather large. Several signs pointed towards an {overestimated} mass and $R_{\rm BLR}$. Namely, it existed far above the $R_{\rm BLR}-L$ relationship when measured by \citet{bentz2006a} and the $M_{\rm BH}-\sigma_*$ relationship measured by \citet{dasyra2007}. Another piece of evidence was that PG~2130+099 exhibits an optical spectrum of a narrow-line Seyfert~1 galaxy, which are supposed to have extremely high accretion rates, but due to its mass measurement at the time had a surprisingly small {accretion} rate instead compared to other narrow-line Seyfert~1 galaxies. 

This prompted further monitoring by \citet{grier2008} explicitly to constrain the radius of the BLR. {They observe with an improved average observational {cadence of 4.9~days (median cadence of 1~day) and,}} despite lower variability from the AGN, they find that cross-correlation between the 5100$\Angstrom$ continuum and H$\beta$ results in a time lag of $22.9^{+4.4}_{4.3}$~days with a SMBH mass of $(3.8\pm1.5) \times 10^7 M_\odot$. These results bring PG~2130+099 back into agreement with both the $R_{\rm BLR}-L$ and $M_{\rm BH}-\sigma_*$ relationships. The reason for previous spurious results may have come from the equivalent width of the {broad} emission lines changing without a {significant change in the measured time lag}. {This raised the total measured emission line flux without an apparent corresponding rise in the continuum. This was a potential indicator that PG~2130+099's BLR was experiencing flux changes unrelated to RM and skewing results.} 

{However, the uneven observational cadence and limited campaign duration of $\sim100$~days prompted further observations to confirm this result.} This object was analyzed again with higher cadence observations by \citet{grier2012}, {finding} that PG~2130+099 had an even smaller {H$\beta$} lag with {$\tau = 12.8^{+1.2}_{-0.9}$~days and resulting mass of $4.6\pm0.4\times10^7M_\odot$}. {While the recovered SMBH masses of both campaigns agree within uncertainty, the new set of lags pushed PG~2130+099 slightly further away from agreement with the $R_{\rm BLR}-L$ relation.} Analyzing the results with velocity-delay maps, \citet{grier2013} finds that, while noisy, {an average} delay of $\sim$30~days is found. If this is accurate, it would lead to a new mass measurement of $\sim10^8M_\odot$ and bring it more in line with the previous relations. 

{A more recent BLR RM measurement comes from another single target study by \citet{hu2020}, who perform a two year high cadence observation campaign from June 2017 to January 2019. Monitoring} H$\beta$, \ion{He}{1}, and {optical} \ion{Fe}{2}, they find the lags for the latter two emission lines change dramatically between years. In the first year the lines show a more typical Keplerian kinematic stratified BLR, i.e. $\tau_{\textrm{He I}} < \tau_{\textrm{H}\beta} < \tau_{\textrm{Fe II}}$. However, the second year shows instead $\tau_{\textrm{He I}} \leq \tau_{\textrm{H}\beta} > \tau_{\textrm{Fe II}}$, potentially indicating a {change has occurred in the structure} faster than the dynamical timescale would allow. This change also coincides with seasonal $F_{\rm 5100}$ variations of $\sim20\%$.

A follow up study by \citet{yao2024} adds four more years of observation from June 2019 to January 2023. Combined with \citet{hu2020}'s data, this creates a six-year observation cycle for PG~2130+099. Their results show again dramatic changes in PG~2130+099's lags, with the two final years of observation resulting in H$\beta$ lags around 60~days, around two times larger than was observed in the previous four years. However, the 5100$\Angstrom$~luminosity does not vary nearly as much, with changes ranging around 10\%. They suggest that a phenomenon called ``geometric dilution" \citep{goad2014} can explain the changes in implied BLR size, where rapid variations in the continuum are coherent in the inner region of the BLR but are diluted further away. During periods with less rapid variability, the {variable emitting region} in the BLR shifts further out and gives a more accurate estimation of the true BLR size. Their lags also imply the BLR changes from inflow to virial {motion} on timescales of a year, alternating back and forth {over} the span of the six years of observation. {These measurements would be in conflict with the estimated dynamical timescale of $\sim$10~years.}

The accretion disk {size estimates suffer similar disagreements between studies. Interband continuum {RM} studies of PG~2130+099 have not been as common, but two previous studies have overlapping {campaigns}, allowing for a more direct comparison of results. The first study by \citet{jha2022} used Zwicky Transient Facility data to study 19 AGN in the $g$, $r$, and $i$ SDSS bands, including PG~2130+099. A notable conclusion from this study is that PG~2130+099 is among 5 sources {for which measured continuum interband delays agree} with thin disk predictions for disk size, which is a rare occurrence for AGN. There are no apparent connections between PG~2130+099 and the other objects in this subset that indicate why {the inferred disk size should} agree with thin disk predictions. However, it should be noted that the observational cadence for the $i$-band data is sparser compared to the $g$- and $r$-band. The other study by \citet{fian2022} observed PG~2130+099 over a 6~month observation campaign from June-December~2019 using 4~narrowband filters to avoid contamination from broad emission lines. These filters had central wavelengths of 4250$\Angstrom$, 5975$\Angstrom$, 7320$\Angstrom$, and 8025$\Angstrom$~and {the monitoring achieved} almost a daily cadence. Using 6~{independent}~lag~determination methods, they find that the disk size {is on average} $\sim$3.7 times larger than predicted {by the standard model}. Of note, both studies observed PG~2130+099 at around the same time. {Jha et al.'s study begins about 60~days earlier and ends about 10~days after Fian et al.'s observations.}

Another study was recently carried out by Miller et al. (in prep.). This study observed PG~2130+099 in addition to 17 other AGN with the Sloan $ugri$ and Pan-STARRS $z_s$ bands for two  observation seasons, 2020 and 2021. Between the two years, two separate lag measurements are recovered, which highlight a change in the state of PG~2130+099. {The $u$-band lags change by a factor of 10 between years, going from consistent with no lag in the 2020 observations to firmly negative with respect to the $g$ band in the 2021 observations. A noted $i$-band excess attributed to H$\alpha$ emission is detected across both years as well.} The interband lags are in line with what was observed by \citet{fian2022}, although we note a large observational gap in the middle of the 2021 observing season that may influence these results. 

{It is worth mentioning that previous continuum observational campaigns of PG~2130+099 as well as many others view the lags through the interpretation of a lamppost model reverberation signal on a Shakura-Sunyaev disk {\citep{1973A&A....24..337S}}. However, there exist alternative explanations for the continuum lags that deserve discussion, especially with the number of AGN that deviate from the expectations of a Shakura-Sunyaev disk. As mentioned above, any gas that produces emission lines will also produce associated continuum emission. This BLR continuum will be present throughout the spectrum, though is especially prominent around the Balmer and Paschen jumps. Continuum emission from the BLR will lengthen the lags {relative to that from the disk alone}, and may even dominate the {delay} signal \citep{korista2001,korista2019,lawther2018,netzer2022}.  Evidence for BLR continuum lags is seen observationally from enhanced lags around the Balmer jump \citep[e.g.,][]{fausnaugh2016,cackett2018}, and Fourier-resolved lags showing the lags are dominated {by the BLR} on long timescales \citep{cackett22,Lewin2023}. {Reprocessed continuum contributions are also apparent in the flux spectrum, the small blue bump - a combination of mostly free-bound Balmer continuum emission and \ion{Fe}{2}}. At high accretion rates, the structure of the disk is expected to change to a ``slim" disk solution \citep{narayan1998, wang1999}. This is of particular interest to PG~2130+099 given that {in a recent campaign PG~2130+099 was observed in such a state \citep{hu2020}.} In this accretion regime, the relationship between observed wavelength and lag changes to $\tau_{\lambda} \propto \lambda^2$. This is due to changes in the geometry for a slim disk, where the inner regions of the disk will increase in height and potentially cast a shadow onto the disk \citep{wang2014}.}

{Other explanations for {a} larger than expected continuum emitting region may be additional factors influencing measurements of the region and artificially increasing the implied size. By accounting for factors such as disk winds and color correction effects, predicted accretion disk sizes can be brought to parity with those observed \citep{zdziarski2022}. Analytic prescriptions that include factors such as corona height, X-ray luminosity, and {SMBH} spin can also match observed {time lags} \citep{kammoun21a, kammoun2023}. The corona-heated accretion disk reprocessing model (CHAR) posits that the outer disk is linked to the corona by the magnetic field, producing coherent temperature fluctuations throughout the disk \citep{sun2020a, sun2020b}. {These alternative models can all explain the interband delays, but additionally need to provide a method for how broad emission lines are produced without a diffuse continuum component.}

To further investigate the variable continuum lags in PG~2130+099, we proposed another study with high cadence observations. To avoid low cadence due to instrument issues or weather, we {monitored with} several remote robotic observatories. Together, these facilities show excellent variability from PG~2130+099 across the widest wavelength coverage and highest cadence from any previous continuum RM study of this source.  

This paper is structured as follows. In Section~\ref{sec:methods}, we discuss the new observations acquired for this study. In Section~\ref{sec:analysis}, we analyze {the} data, and determine interband lags using several different methods. In Section~\ref{sec:discussion} {we discuss the results within the wider context of previous results}, and in Section~\ref{sec:conclusion} we summarize and review our findings. 

\section{Methods}
\label{sec:methods}
Observations of PG~2130+099 occurred from July 21st, 2023 to January 9th, 2024, spanning almost 6~months. These observations were taken with the Dan Zowada Memorial Observatory (hereafter Zowada), the Liverpool Telescope (LT) and several {telescopes from} the Las Cumbres Observatory (LCO) network. {The LCO} telescopes are referred to by their designation within the LCO network, and {are} all 1~m observatories.

The Zowada observatory \citep{carr2022}, located in Rodeo, New Mexico, is a 0.5m telescope owned and operated by Wayne State University. It observes with the $ugri$ SDSS and Pan-STARRs $z_s$ (hereafter just $z$) filters. LT, a 2m telescope, is located in the Canary Islands \citep{2004SPIE.5489..679S}. It contains the same filters, save for the SDSS $z$ band instead of the Pan-STARRs $z_s$. Finally, we use several observatories that are a part of the LCO network, a global array of robotic observatories located at many different facilities \citep{2013PASP..125.1031B}. Each are equipped with the same filters as Zowada. The full list of observations are given in Table~\ref{table:telescope_info}.

\input{lightcurve_table}

We obtain light curves using differential {broadband} photometry. We select comparison stars to maximize signal to noise ratio of our data, although our selection is limited to stars present in the field of view shared by all observatories. We also select different comparison stars for the $u$-band due to many stars being much dimmer in this particular band. We find the AGN and stars in our images using the photutils \citep{photutils2021} module DAOStarFinder {in Python}. Once identified, we {use a circular} aperture and annuli to determine the source brightness and background. Background annuli with an inner and outer radius of 20~pixels and 30~pixels respectively are used for the Zowada observations and scaled to match the same angular size for the other observatories. The sizes for the source apertures are taken from \citet{Miller2023} as that study uses the same observatories and performed testing to maximize the signal-to-noise ratio of observations. For Zowada this is an aperture radius of 5~pixels and we use 8-pixels for LT. We attempt a similar test of aperture sizes for the LCO observatories not included in that study but find no consensus on a size for each facility, so we elect to use the previously found value for the 1m observatories of 11~pixels. We use the same stars for all observatories except for the Zowada $u$-band observations, which uses a different set of stars to maximize the signal to noise. 

{For each observatory, we perform sigma clipping on the comparison star light curves, removing observations when comparison stars deviate from their respective means by more than $3\sigma$. For the AGN, each observation is subtracted by a moving boxcar average. If this value is greater than the error of the observation multiplied by a scale factor, that observation is removed. The scale factor is typically between 5 and 20 and is scaled manually depending on the quality of the light curves. This is done to remove extreme variations in the light curve uncharacteristic of Type 1 Seyferts that are likely due to poor observing conditions. In addition, we inspect images by eye to confirm that no cosmic rays are near the object, and remove any images from our final sample that are affected.}

{We use CALI \citep{2014ApJ...786L...6L} to combine the light curves from each observatory.} CALI adjusts the data by shifting and scaling light curves to a common {flux}~scale. It assumes that the variability is modeled by a {damped} random walk {(DRW)}, creating a model of the light curve and using it to interpolate between any observing gaps. Additional systematic uncertainties are added in quadrature to the data to account for the differences between these facilities. For this process, we {set} the error scaling functionality {to allow} only the systematic error term from CALI to account for the additional systematic uncertainty.

Finally, we convert the relative fluxes to physical fluxes using comparison stars available with measured fluxes in stellar catalogs. For the $g$, $r$, and $i$ bands we use APASS data release 10 \citep{henden2018} and for the $u$ and $z$ bands we use SDSS data release 18 \citep{almeida2023}. We perform this conversion using the Zowada data as reference as they are used as the reference light curve by CALI and not scaled or shifted as a result. We find all stellar magnitudes in the Zowada's field of view from the first night of observations and use these to {set the flux scale.} The only exception to this process is for the $z$-band data. Unfortunately, Zowada data were too noisy to be used in the intercalibration process, and further analyses suffered as a result. {When Zowada is excluded and instead LT is used as the reference band for intercalibration, we are able to resolve the peaks and troughs of the final light curve, which are required for robust RM analyzes.} As such, we exclude the Zowada $z$-band data {from the analysis} and use LT for all of the above processes for flux conversion in the $z$-band. These fluxes are presented in Fig.~\ref{fig:light curves}. {We recover a fractional variability {amplitude \citep{edelson1990,rodriguez-pascual1997, vaughan2003}} of $5.2\pm0.3\%, 3.7\pm0.1\%, 3.2\pm0.1\%, 2.4\pm0.1\%,$ and $2.0\pm 0.1\%$ for the $ugriz$ bands, respectively .} 

\begin{figure*}
 \centering
 \includegraphics[width=0.99\textwidth]{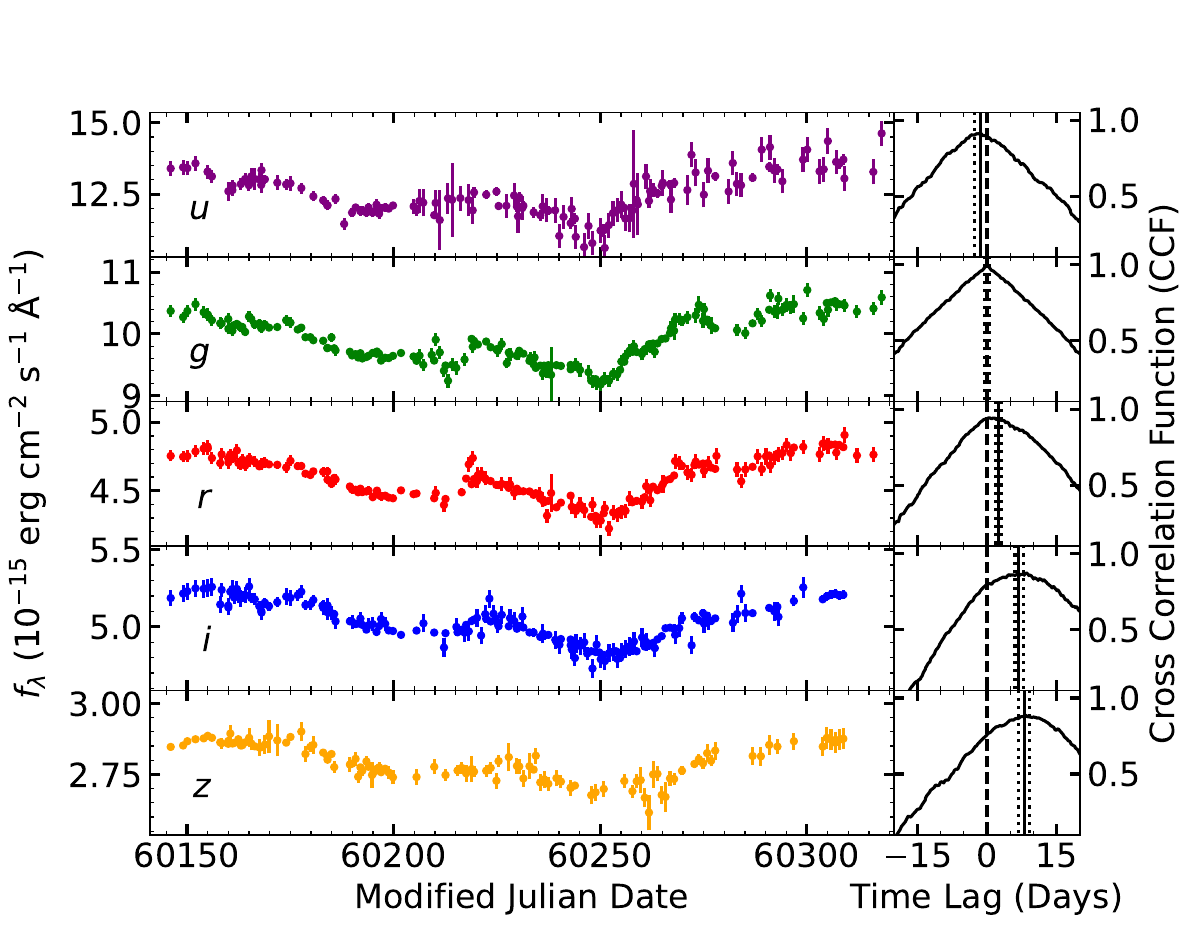}
 \caption{Light curves of PG~2130+099 in the $ugriz$ bands. The panels on the right are the CCF of each band with respect to the $g$ band. The solid vertical line is the measured lag {from PyCCF}, with the dotted lines representing the uncertainty on the lag and the dashed line indicating a lag of 0 days for reference.}
 \label{fig:light curves}
 \end{figure*}

\section{Analysis}
\label{sec:analysis}
\subsection{Time Lags}
We measure the time lags between the light curves using {3 methods:} the interpolated cross-correlation function (ICCF) \citet{koratkar1991} method as {implemented} by \citet{peterson2004}, {a DRW} model, and a running optimal average (ROA). To do this, we use the modules PyCCF \citep{sun2018}, JAVELIN \citep{zu2013}, and PyROA \citep{donnan2021} {with Python 3}.

PyCCF uses flux randomization and random subset sampling {\citep{peterson1998}} to create different realizations of the light curves. Using these realizations, it linearly interpolates between the data points in the light curves and calculates the CCF between them. To determine the lag, it calculates the centroid for all values above 80\% of the CCF's peak. Finally, it repeats this process 10,000 times to create a distribution of CCF {centroids}, of which the median is taken as the lag and its uncertainties are taken at 16\% and 84\% of the distribution. We attempt to find lags using a lag range of $-30$ to 40~days, which is determined from observing the shape of the CCF. 

An advantage of the CCF method is that it is model agnostic and contains no assumptions about the AGN variability. A downside of this is that typically PyCCF reports larger uncertainties in the lags \citep{yu2020b}. An alternative approach is {to} model the light curves, assuming that all the light curves are smoothed and delayed versions of each other. By assuming a model for the variability a more realistic interpolation between the data points can be {performed}. A common model for AGN variability is a {DRW}. JAVELIN (Just Another Vehicle for Estimating Lags In Nuclei) {fits} a model for the driving light curve, assuming the light curves in all other bands are just lagged and smoothed in relation to it. PyROA models AGN variability with a ROA, and as such does not assume a shape for the variability. It instead creates the model from the data itself using a Gaussian window function. Once the driving light curve is created, it is then scaled and shifted to match the data in all the bands, thus measuring the time lags. 

{For both JAVELIN and PyROA, we first measure the lags of all light curves simultaneously using the $g$ band as the reference band. We then measure only the $g$ band against itself. This is done to prevent the $g$ band from being double-counted in the multi-band analysis while still measuring a lag and uncertainty.} For both methods we restrict the lag ranges to be between $-10$ and $20$~days. {The light curve models and prior distributions for PyROA and JAVELIN can be found in the Appendix.} All lags measured from the above methods are plotted in the rest frame {assuming a redshift $z=0.06298$ in} Fig.~\ref{fig:timelags}} and are given in Table~\ref{table:lags}.

\input{Lagtable}

Assuming an optically-thick and geometrically-thin accretion disk \citep{1973A&A....24..337S}, the disk's temperature~($T$) will depend on its radius~($R$), {SMBH} mass~($M$), and accretion~rate~($\dot{M}$) following $T(R) \propto (M\dot{M})^{1/4}R^{-3/4}$. The disk is expected to {emit} like a blackbody, and thus we can {use Wien's law $\lambda_{\textrm{Maximum}}$ $\propto$ $T^{-1}$} to exchange temperature for wavelength~$\lambda$. Assuming the X-ray corona emits isotropically and that this powers the optical variability on short timescales, we can then use geometry to determine the time lags. The time lag ($\tau$) would then simply be the light travel time, $\tau \sim R/c$. We can combine the above equations to relate the lag to emitted wavelength {for the irradiated thin disk model:}
\begin{equation}
\tau(\lambda) \propto (M\dot{M})^{1/3}T^{-4/3} \propto (M\dot{M})^{1/3}\lambda^{4/3}.
\label{eq:lageq}
\end{equation}

Previous {RM} campaigns have found broad agreement with this prediction, despite purported interference from other regions. 
{In particular, the BLR can contribute slowly varying optical emission through the Balmer and Paschen continua and broad-emission lines \citep{korista2001, korista2019}. These emissions could interfere with lag determinations, as they vary on much longer timescales ($\sim$weeks-months).} {Reprocessed continuum from the BLR} is used to explain the excess lags observed in the $u$-band and around the Balmer jump \citep[e.g.,][]{fausnaugh2016, cackett2018, edelson2019}}, but will also contribute at all wavelengths and approximately mimics $\tau(\lambda) \propto \lambda^{4/3}$ {through the Paschen continuum} \citep{korista2001}. Some studies report that the variability observed within continuum RM studies is dominated by BLR continuum emission \citep{chelouche2019, netzer2022, guo2022, montano2022}, although a consensus on this has not been reached. If true, then it could explain the common observation that {inferred} disk radii are larger than expected compared to theory. {However, this disparity could also be rectified through other methods, such as including internal reddening \citep{gaskell2018} or by considering different assumptions when calculating disk size \citep{2018MNRAS.473...80T}.}

\begin{figure}
 \centering
 \includegraphics[width=0.5\textwidth]{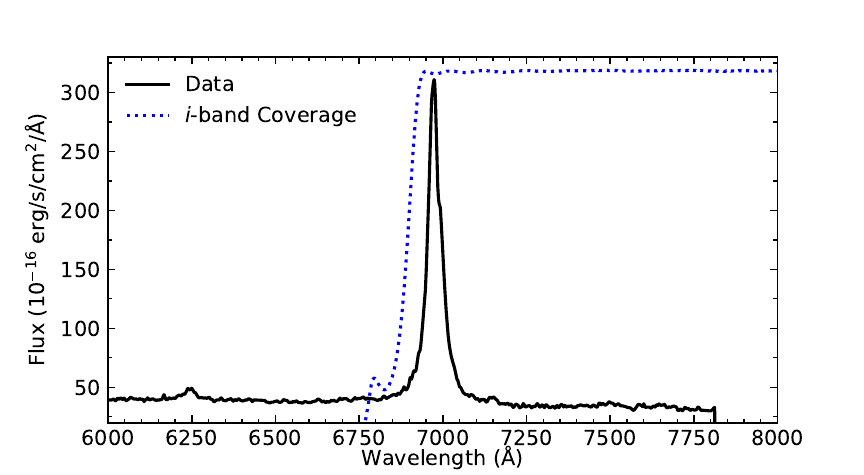}
 \caption{{An observed spectrum of PG~2130+099 taken in September of 1998 as a part of \citet{2000ApJ...533..631K}'s observations. The spectrum is presented in the observed frame as the black solid line, and the wavelength range observed by the $i$ band is shown in the dotted blue line. The redshift of PG~2130+099 shifts the H$\alpha$~emission line into the $i$-band's coverage, and likely adds a non-negligible contribution to the observed lag.}}
 \label{fig:spectra}
 \end{figure}

{We first compare our measured lags to the disk reprocessing model.} To {fit the lags}, we use the following {relation}
\begin{equation}
\tau(\lambda) = \tau_0[(\lambda/\lambda_0)^\beta - 1],
\label{eq:wavlag}
\end{equation}
where $\tau$ is the measured time lag, $\tau_0$ is the {lag} at the chosen reference band, {and} $\lambda_0$ the wavelength of the reference band. $\beta$ is assumed to be 4/3 if the {lags follow expectations for a} Shakura-Sunyaev accretion disk. We plot the fits of Eq. \ref{eq:wavlag} in Fig.~\ref{fig:timelags} with $\beta$ fixed to be 4/3 (solid black line) and allowing it to be free (dashed {yellow} line). We find {reasonable} agreement with the 4/3 power predicted by a Shakura-Sunyaev disk across all three lag determination methods. We find the presence of an excess $i$-band lag, especially with the JAVELIN and PyROA methods. This is likely due to the presence of H$\alpha$ emission line, which given PG~2130+099's redshift would affect the $i$-band lags specifically. {This phenomenon has been observed in other objects, as reported by \citet{Miller2023}. Notably, \citet{fian2022} employed special narrowband filters in their study of PG~2130+099 to mitigate potential contamination by broad emission lines.} {This is visually shown by plotting the observed spectra obtained by \citet{2000ApJ...533..631K} on top of the $i$-band wavelength coverage, which is presented in Fig.~\ref{fig:spectra}.}

{To avoid potential H$\alpha$ contamination in the continuum emitting region size measurement, we}  fit the lags excluding the $i$-band lag (red lines). We find a much better fit with this exclusion, with all methods more closely following the $\beta$~=~$4/3$ prediction for a Shakura-Sunyaev disk.
{When excluding the $i$-band, the $\tau_0$ measurements are reduced and have smaller uncertainties.~This leads us to believe that the H$\alpha$ contribution presented in Fig.~\ref{fig:spectra} is significant enough to influence the disk size measurement.}~As such, we will use {the $\tau_0$ found when excluding the $i$-band} going forward with our analysis. The lags and $\tau_0$ values both with and without the $i$-band lag are presented in Table~\ref{table:lags}. {With the $i$ band excluded, we find the best fitting $\beta$ {(dashed yellow line)} of all methods are consistent with a slim disk interpretation {(solid blue line)}, but can also be explained with the standard thin disk interpretation as well}.  
{We also include the lags predicted for various values of $X$, which are calculated in the next section. We cannot model the analytic prescription of \citet{kammoun2023} as we do not have simultaneous X-ray observations, which are necessary to constrain the fit.}

\begin{figure*}
 \centering
 \includegraphics[width=0.99\textwidth, trim={1.8cm 0 2cm 0},clip]{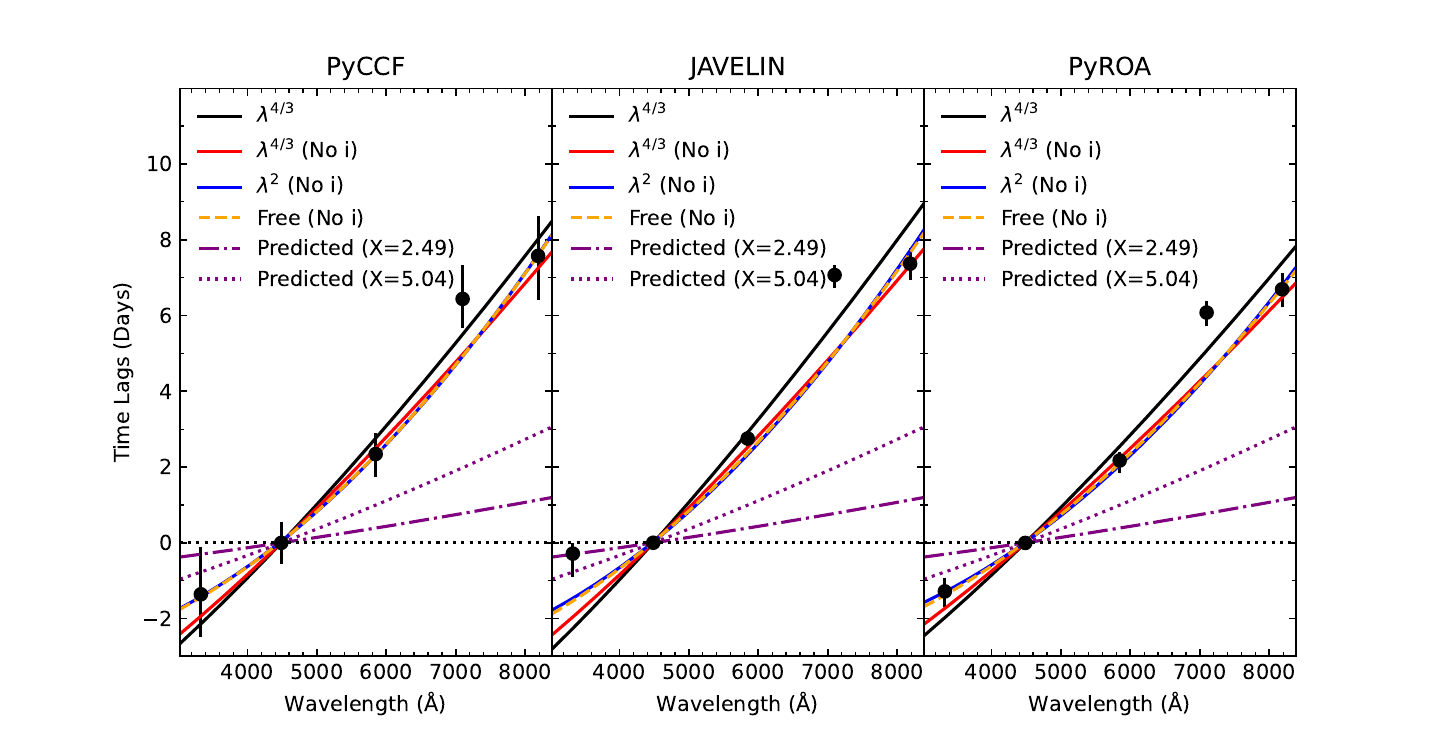}
 \caption{{Measured lags for PG~2130+099 using PyCCF (left), JAVELIN (middle) and PyROA (right). We initially fit the lags using Eq.~\ref{eq:wavlag} with $\beta$ fixed to 4/3 (solid black line). However, the $i$-band {lag (point at $\sim$$7000\Angstrom$)} clearly shows deviation from the expected trend, so we exclude it from the rest of the fits. We fit excluding the $i$-band lag for $\beta$ fixed at 4/3 (red line), at 2 (blue line), and as a free parameter (dashed {yellow} line). We also plot the expected time delays calculated in Section~\ref{sec:theoretical_disk_size} for both $X=2.49$ (dashdot purple line) and $X=5.04$ (dotted purple line), corresponding to $\tau_0$ = 0.92~days and 2.35~days respectively.}}
 \label{fig:timelags}
 \end{figure*}

\subsection{Theoretical Disk Size Comparison}
\label{sec:theoretical_disk_size}
We compare the measured {lags} to the predicted disk size following the parameterization of \citet{fausnaugh2016},
\begin{equation}
\tau_0 = \frac{1}{c} \left(X\frac{k \lambda_0}{hc}\right)^{4/3} \left[\left(\frac{GM}{8 \pi \sigma}\right) \left(\frac{L_{\mathrm{Edd}}}{\eta c^2}\right) \left( 3+\kappa \right) \dot{m}_{\rm E} \right]^{1/3},
\label{eq:alpha}
\end{equation}
where the {lag} ($\tau_0$) {at wavelength $\lambda_0$ depends} on a few parameters such as the Eddington luminosity ($L_{Edd}$), Eddington fraction ($\dot{m}_{\rm E}$), and a unitless parameter $X$ which is used to convert temperature to wavelength at a given radius. {A range of disk radii contribute to the total emission at any given wavelength. This factor is used to account for this additional inclusion. The calculation of $X$ requires X-ray observations to constrain the temperature profile of the disk, which we do not have available for this study.} \citet{fausnaugh2016} {derive} a value of $X$=2.49. Other studies have since recalculated $X$ {with consideration of additional effects from the disk. One such factor is the variability of the disk emission. The length of time it takes for disk emission to vary lengthens with distance from the SMBH. This variability will extend the possible regions for which any wavelength can be produced. With this inclusion, $X$ increases to 5.04 \citep{2018MNRAS.473...80T}.} We assume that the radiative efficiency $\eta=0.1$ and that the ratio of external to internal heating $\kappa=1$. We use the mass as listed in the AGN Black Hole Mass Database \citep{bentz2015} of {M~$=2.71\times10^7\mathrm{M_\odot}$}. {We interpolate the measured $g$ and $r$-band fluxes to approximate the luminosity at 5100$\Angstrom$ {, which we find to be $3.48\times10^{44} \pm1.87\times10^{43}$ erg/s We then subtract a host galaxy luminosity contribution of $3.30\times10^{43}\pm1.65\times10^{42}$} erg s$^{-1}$ found from HST modeling \citep{bentz2009, bentz2013} {to determine an AGN luminosity at 5100$\Angstrom$ of $3.12\times10^{44}\pm1.87\times10^{43}$ erg s$^{-1}$}.  We use this to} estimate the Eddington fraction using a bolometric correction of $L_{\rm bol} \sim 9\lambda L_{\lambda}(5100\angstrom)$ \citep{2000ApJ...533..631K}.

{{When using our calculated value of $\dot{m}_{\rm E}=0.91$ and a value of $X=2.49$, we find a predicted size for the accretion disk to be $\tau_0=0.91$~days. {Our measured values for $\tau_0$ are 6.48, 6.55, and 5.79}} times larger for PyCCF, JAVELIN and PyROA respectively. When using a larger value for $X$ such as 5.04, we calculate a disk size of $\tau_0=2.33$~days, with our measured sizes now  2.53, 2.56, and 2.26 times larger for PyCCF, JAVELIN, and PyROA respectively.} 

While a larger value of $X$ brings the theoretical disk sizes closer to agreement with observations, there is still a factor of $\sim2$ discrepancy between them. It may be that additional factors not yet accounted for in $X$'s determination are required. Alternatively, it may be that there are contaminating sources (i.e. the BLR's diffuse continuum) that lengthen the reverberation measurements of continuum emitting regions.~From our analysis, it is impossible to determine which is the correct interpretation. It is likely a combination of these two factors that will lead to agreement between observation and theory.  Moreover, disk models that properly account for {general relativity} ~and other effects have been shown to fit continuum lags well \citep{kammoun21b,kammoun21a,kammoun2023}. However, since we do not have simultaneous X-ray observations we cannot test these more complicated interpretations.

\subsection{Flux-Flux Analysis}
\label{sec:fluxflux_analysis}
\begin{figure*}
 \centering
 \includegraphics[width=0.99\textwidth]{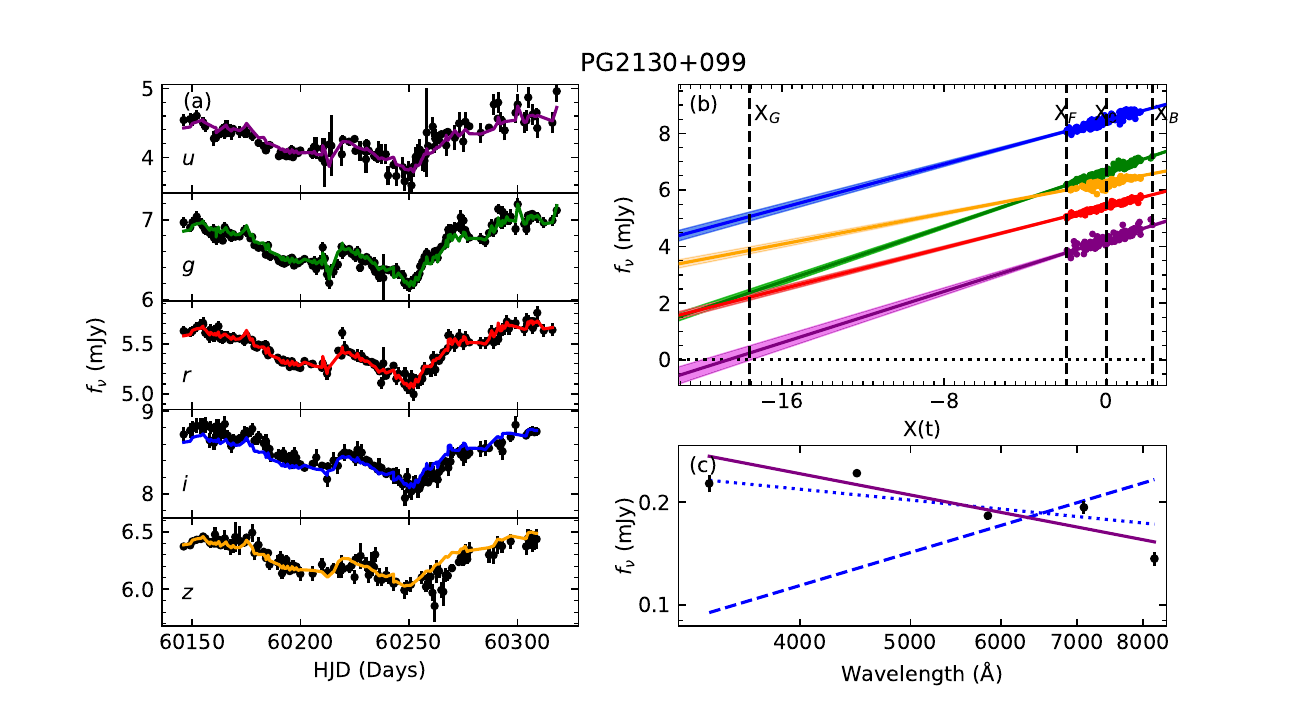}
 \caption{{Flux-flux} analysis for PG~2130+099. Panel (a) presents the fluxes in units of mJy and the fit of the flux-flux analysis (Eq.~\ref{eq:fluxflux}) to the light curve. Each filter's color is carried over to panel~(b), which plots the values of the flux against the fitted value of X(t). Panel~(c) plots the RMS {flux} against wavelength, with the lines representing powerlaw fits to the data. The dotted blue line is the {$\lambda^{-1/3}$ relation expected} for a Shakura-Sunyaev disk, {the dashed blue line is the {$\lambda^{+1}$ relation expected for a} super-Eddington slim disk \citep{wang1999, donnan2023}, and} the solid purple line allowed the exponent to be a free parameter. This freely {fitted exponent is found to be $-0.64 +/- 0.22$}.  An analysis of these results is presented in Sec.~\ref{sec:fluxflux_analysis}.}
 \label{fig:fluxflux}
 \end{figure*}

We can split the variable (AGN) and constant (host galaxy) flux using the flux-flux analysis \citep[e.g.,][]{mchardy18, cackett2020, Miller2023}. This {method} models the light curves as a combination of these two components, parameterizing it as 
\begin{equation}
f_{\nu}(\lambda, t) = A_{\nu}(\lambda) + R_{\nu}(\lambda) X(t),
\label{eq:fluxflux}
\end{equation}
where our fluxes ($f_\nu$) are equal to the average spectrum ($A_\nu$), representing the constant flux, added to the {driving light curve $X(t)$ multiplied by} the rms spectrum $R_\nu$. $X(t)$ is normalized to have a mean of 0 and a standard deviation of 1, such that $A_\nu$ and $R_\nu$ can be used to scale $X(t)$ to each light curve. $R_\nu$ represents the variable component, and for a Shakura-Sunyaev thin disk is expected to scale with the flux like $f_\nu \propto \lambda^{-1/3}$. {However, for a slim disk, the scaling relation changes considerably, such that $f_\nu \propto \lambda^{1}$ \citep{wang1999, donnan2023}.} This analysis is shown in Fig.~\ref{fig:fluxflux}, and values are presented in Table~\ref{table:fluxflux}. Panel (a) presents the fluxes with the fit of Eq~\ref{eq:fluxflux} overlaid. Panel (b) plots the fluxes (in units of mJy) against $X(t)$ which is used to determine the constant contribution of the fluxes. {The uncertainties for each line are shown as the colored envelopes.} The $u$ band is used as reference, with its lower bound on the uncertainty {taken} as the {minimum constant} value, allowing it to have a constant measure above 0 as well. It is expected that longer wavelengths have a larger contribution from the host galaxy, and in general that is seen. However, the band with the largest constant contribution is the $i$ band, indicating the presence of an additional component in our observations. Panel (c) plots the RMS values found from the fit against wavelength. For a Shakura-Sunyaev disk, we would expect these points to follow $\lambda^{-1/3}$ power law relation. We fit {fixing the exponent to be $-1/3$ (blue dotted line), fixing the exponent to be $+1$ (blue dashed line), and allowing it to be a free parameter (solid purple line). {The free-fitted exponent is {found to be $-0.64 \pm 0.22$,} deviating from what is expected for a simple thin disk.} This is consistent with a steeper radial profile predicted for relativistic disks with innermost stable circular orbit stresses \citep{mummery2020}. This is also consistent with the average result found from similar analyses on SDSS quasars \citep{weaver2022}. This model predicts a steeper temperature relation than the standard thin disk of {$T(R) \propto R^{-7/8}$}, which in turn predicts $f_\nu \propto \lambda^{-5/7}$, consistent with what we observe in PG~2130+099.} {Notably, we do detect a slight $i$-band RMS excess, which was also found in a previous study of Mrk~876 where H$\alpha$ was determined to contribute significantly to an $i$-band lag excess \citep{Miller2023}}

\input{FluxFlux_Fnu_PG2130+099__ugriz_Values}

\section{Discussion}
\label{sec:discussion}
Over roughly 6~months, we obtained {observations with a median cadence of 2~days in the $g$-band} of PG~2130+099 during which it was significantly variable. We recover robust time lags between the $g$ band and the $u$, $r$, $i$ and $z$ band light curves with each lag method (PyCCF, JAVELIN, PyROA). {The lags are all consistent with each other and generally follow the $\lambda^{4/3}$ expectation for a Shakura-Sunyaev disk. The only major difference between all the methods are the larger uncertainties recovered with PyCCF. This is not surprising, as multiple other studies have found similarly that PyCCF-determined lags {tend to have larger uncertainties \citep{edelson2019, Homayouni2019, guo2022}}. }~We find an $i$-band excess lag with all methods. Removing this lag results in a significantly better fit to $\lambda^{4/3}$, and results in a value of $\tau_0$ that is at least 1$\sigma$ different and {with} smaller uncertainties than with the $i$-band included. This lag excess is likely due to H$\alpha$ emission. {The $i$-band bandpass ranges from $6700-8700\Angstrom$ with a pivot wavelength of $7718.28\Angstrom$.} At the redshift of PG~2130+099 this would {shift the emission line to 6978$\Angstrom$ and thus into} the $i$-band. As such, the $i$-band lag is excluded and this value of $\tau_0$ is used for all further analysis unless otherwise stated. {With the $i$-band lags excluded, we find that the best fitting $\beta$ for each method can be fit sufficiently by either the expectation for a slim disk ($\beta=2$) or a thin disk ($\beta=4/3$).} The continuum reprocessing size determined by PyCCF disagrees with the size found by PYROA, while both agree with the size found by JAVELIN within uncertainties. 

We perform a flux-flux analysis, finding an excess in both the constant and variable emission in the $i$-band. This same excess is seen across all three lag methods in Fig.~\ref{fig:timelags}. We note that the $i$-band model light curve fit is poor, and as a result may not reflect the true RMS present for the duration of the campaign.
It may be that the driving light curve model utilized by the flux-flux method (Eq.~\ref{eq:fluxflux}) is unable to completely reproduce the observed variability. This may be due to another varying component contributing to the observed flux, in particular H$\alpha$.
In addition, we note a slight $g$-band excess detected in the constant and variable emission, which could indicate that the H$\beta$ emission line is contributing to the flux in this band. Flux-flux analysis techniques have recently been shown to be unable to reliably estimate the host galaxy (constant) contribution \citep{netzer2024, cai2024}, so the enhanced $g$-band and $i$-band emissions may not be indicative of physical properties of PG~2130+099.The RMS flux agrees within uncertainty with predictions for a relativistic disk with finite stresses at the inner radius \citep{mummery2020, weaver2022}. This could imply that turbulent magnetic forces close to the {SMBH} are a significant factor in the overall emission.

\subsection{Comparison with Previous Studies}
\label{sec:comp_studies}
\begin{figure*}
 \centering
 \includegraphics[width=0.85\textwidth]{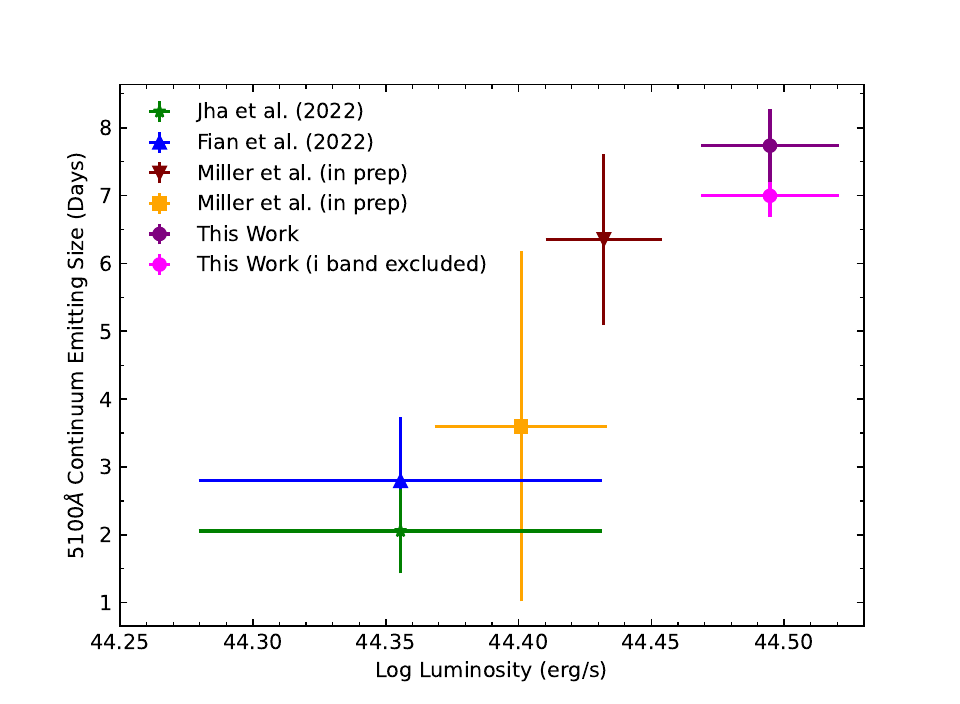}
 \caption{ Comparison of measured ICCF continuum emitting sizes and host-subtracted 5100$\Angstrom$~luminosities for PG~2130+099. The lags from Miller et al. (in prep) are taken from 2020 (brown triangle) and 2021 (yellow square), respectively. Lags are adopted from each study and fit with Eq.~\ref{eq:wavlag} to measure $\tau_0$. This value is then shifted following the fitted $\lambda^{4/3}$ relation to find the implied disk size at $5100$$\Angstrom$.  All values of luminosity are host galaxy subtracted and corrected for extinction and redshift. See Sec. \ref{sec:comp_studies} for details and Table~\ref{table:disksize_lum} for the values of the sizes and luminosities.}
 \label{fig:previous_disk_size_comp}
 \end{figure*}

To directly compare our results to previous studies, we gather the ICCF lags from previous papers and measure {$\tau_0$} using Eq.~\ref{eq:wavlag} in accordance to how the previous studies measured {their $\lambda_0$, then shift $\tau_0$ to a common $\lambda_0$ of 5100$\Angstrom$.} If the studies report the lags in the observed frame, we shift them to rest frame. As Zowada data make up the majority of the light curve for our $g$ and $r$ bands and is used as the baseline for light curve integration, we {subtract} the same value for host galaxy flux as was used in Miller et al. ({in prep.}). Jha et al.'s study does not report a value for 5100$\Angstrom$ flux, so we use the measurement determined by Fian et al. Both studies occur contemporaneously with each other and Fian et al. carefully subtract out host-galaxy contribution to get their measurement. These values are shown in Fig.~\ref{fig:previous_disk_size_comp} and given in Table~\ref{table:disksize_lum}.

\input{DiskSizeTable_Update}

Despite the studies finding different conclusions with respect to agreement of a thin-disk, when normalizing their results we find that \citet{fian2022}'s and \citet{jha2022}'s studies agree within {the} uncertainty for their measured disk sizes. The disparity in their findings originate from several sources. The calculation of the predicted size for a Shakura-Sunyaev disk was performed differently between the two studies. When adjusting for redshift and measuring from the same reference frame, the two studies are brought into agreement with each other, as would be expected for observations of the same object at nearly the same time. {Our calculations of the sizes of the continuum emitting region from their studies disagree with the predicted continuum emitting size we calculate in Sec.~\ref{sec:theoretical_disk_size} of $\tau_0=0.91$~{days} when $X=2.49$, but agree with $\tau_0=2.33$~{days} which is calculated when $X=5.04$ is used}.
 
\begin{figure*}
 \centering
 \includegraphics[width=0.85\textwidth]{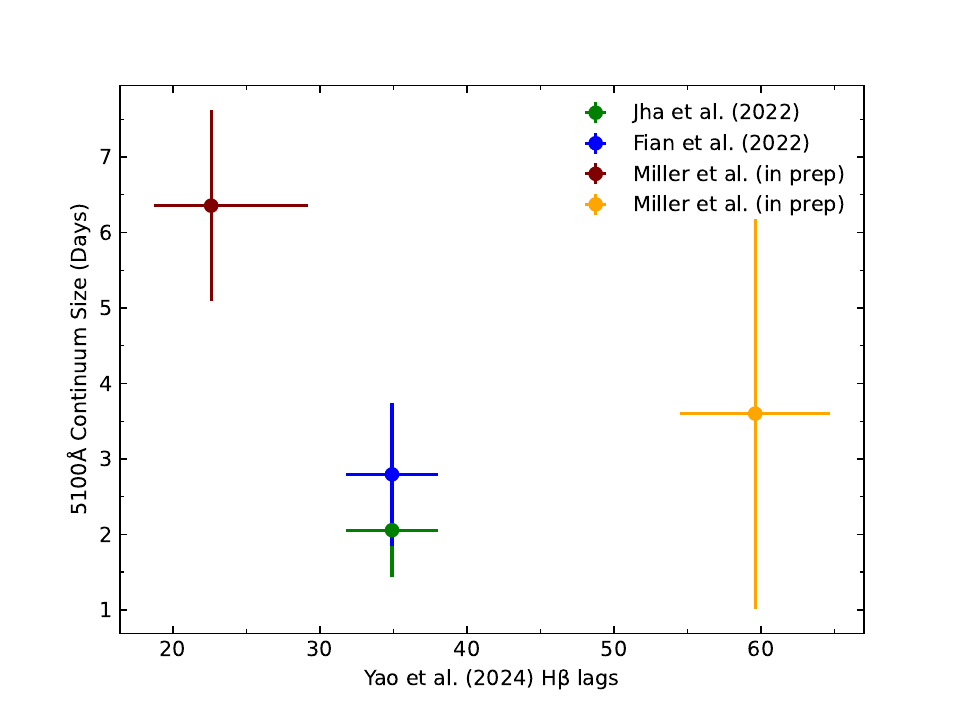}
 \caption{Comparison of continuum emitting regions found by previous studies with the H$\beta$ lags reported by \citet{yao2024}. The two years of observation from Miller et al. (in prep) are obtained during similar time periods to the 2020 and 2021 observations of Yao et al. We see no distinct trends between the sets of lags. We do not find any trend between the two with our current data.}
 \label{fig:continuum_vs_hbeta}
 \end{figure*}

\citet{fian2022} and \citet{jha2022}'s study both correspond to a time identified by \citet{yao2024} when the H$\beta$ line width was at its broadest and among the smaller BLR sizes determined by their campaign. The Miller et al. disk sizes are measured simultaneously with the H$\beta$ measurements reported by \citet{yao2024}. The large shift in the H$\beta$ lag from 2020--2021 is not reflected in the implied continuum reprocessing sizes observed in Miller et al. (in prep) for the 2020 and 2021 (brown triangle and yellow square points, respectively), as shown in Fig.~\ref{fig:continuum_vs_hbeta}. 

\citet{yao2024} also report the inflow/virial motion of the BLR gas for each year of observation, which from 2017-2022 are listed as virial-inflow-inflow-virial-inflow-virial. Jha et al. and Fian et al.'s studies both report disk measurements from 2019, indicated as inflow states. Miller et al. (in prep)'s data from 2020 would then be virial motion, followed by the 2021 data occurring as an inflow. This study's observations are not coincident with the last observations of Yao et al., but appear to follow the trend of longer lags reported when BLR gas exhibits virial motion. However, given the fact that these measurements can change on a yearly basis, this interpretation should be taken with some caution. There is a disconnect when comparing the measured lags. Fian et al.'s and Jha et al.'s lags are the smallest in our set, and align with the smaller BLR radius measured by Hu et al. The H$\beta$ lag stays mostly constant from 2017-2020. This is in contrast to our results, which show more variation from 2019-2020. 

The reported explanation behind smaller implied BLR sizes is given to be geometric dilution \citep{goad2014}. In this scenario, changes in the continuum occur too rapidly for the outer regions of the BLR to respond coherently, leading to a bias towards the inner regions in the reverberation signal. 
The BLR may also contribute significantly to the observed optical continuum \citep{korista2001, korista2019}, and the BLR has been invoked previously to explain larger than expected continuum emitting sizes \citep{fausnaugh2016, cackett2018, lawther2018, netzer2022}. If both of the above scenarios are occurring, then the length of measured continuum lags should also decrease during periods of geometric dilution, {as observed} for H$\beta$. This is tested in Fig.~\ref{fig:continuum_vs_hbeta}. Of the four years of observations from Yao et al., three of them (2019-2021) occur concurrently with continuum reverberation studies. Jha et al. and Fian et al. both observe in 2019, and Miller et al. (in prep) had observations taken in 2020 and 2021.  

Interestingly, we see some evidence that smaller BLR lags correspond to larger continuum lags. We are unable to draw any definitive conclusions from this, as the large uncertainties of the second year of Miller et al. (in prep)'s analysis had an observational gap that lead to large uncertainties in the lag measurements. Additional simultaneous continuum and BLR {RM} studies would be needed to draw any significant conclusions, and more such studies could provide valuable insight into the workings of the inner AGN system.

\section{Conclusion}
\label{sec:conclusion}
We {observed} PG~2130+099 with 10~different telescopes to compile a $\sim$6~month light curve and {measured} lags between photometric bands across the widest wavelength range yet observed {in a continuum {RM} study}. PG~2130+099 appears to be in or near a super-Eddington accretion state, and we {measure the $u$ to $z$ lags of approximately 7~days}. The host-subtracted 5100$\Angstrom$~luminosity is determined and compared to previous studies that {have also} calculated this value. We {report} that the luminosity of PG~2130+099 can change as much as 18\% {from year to year}. 

We ultimately find that {the magnitude of the measured interband delays} disagrees with disk size predictions from Shakura-Sunyaev thin-disk theory.  Our disk sizes are consistent with those reported in Miller et al. (in prep.)'s first year of analysis for PG~2130+099  (2020), but disagree with the continuum-emitting size found during the second year (2021). 

Combining our results with those presented in \citet{yao2024}, we find that the periods of larger/smaller disk sizes tentatively correspond to the periods of inflow/virial motion. This is only tentative evidence, given that Jha et al.'s analysis does not have accurate host galaxy background subtraction and may in reality be slightly different than what is presented here.

This analysis highlights the importance of concurrent studies regarding the accretion disk and BLR, as well as revisiting previously reverberation-mapped objects to determine how their reprocessing regions change over time. More studies, ideally analyzing both the disk and the BLR simultaneously, would prove vital to understanding the connection between these regions in AGN. There is some evidence in this study that the two are linked, and that larger continuum emitting regions are measured when the BLR displays virial motion as opposed to inflow motion.

\bibliography{PG2130}{}
\bibliographystyle{mnras}

\appendix
{We present here the results from the PyROA and JAVELIN modeling that produces the lags shown in Fig.~\ref{fig:timelags} and {presented in} Table~\ref{table:lags}. The light curves displayed here are identical to those in Fig.~\ref{fig:light curves}, with the addition of the model light curves generated by PyROA/JAVELIN overlaid. {The shaded regions overlaid on the models/light curves} represent the $1-\sigma$ uncertainty in the light curve models. The posterior distributions are included on the right, with the measured lags from each distribution marked similarly to those presented in Fig.~\ref{fig:light curves}. The solid black lines are the lag measurements, with the dotted lines on either side representing the uncertainties. The dashed black line is at 0 for reference.
}

\begin{figure*}[h!]
 \centering
 \includegraphics[width=0.99\textwidth]{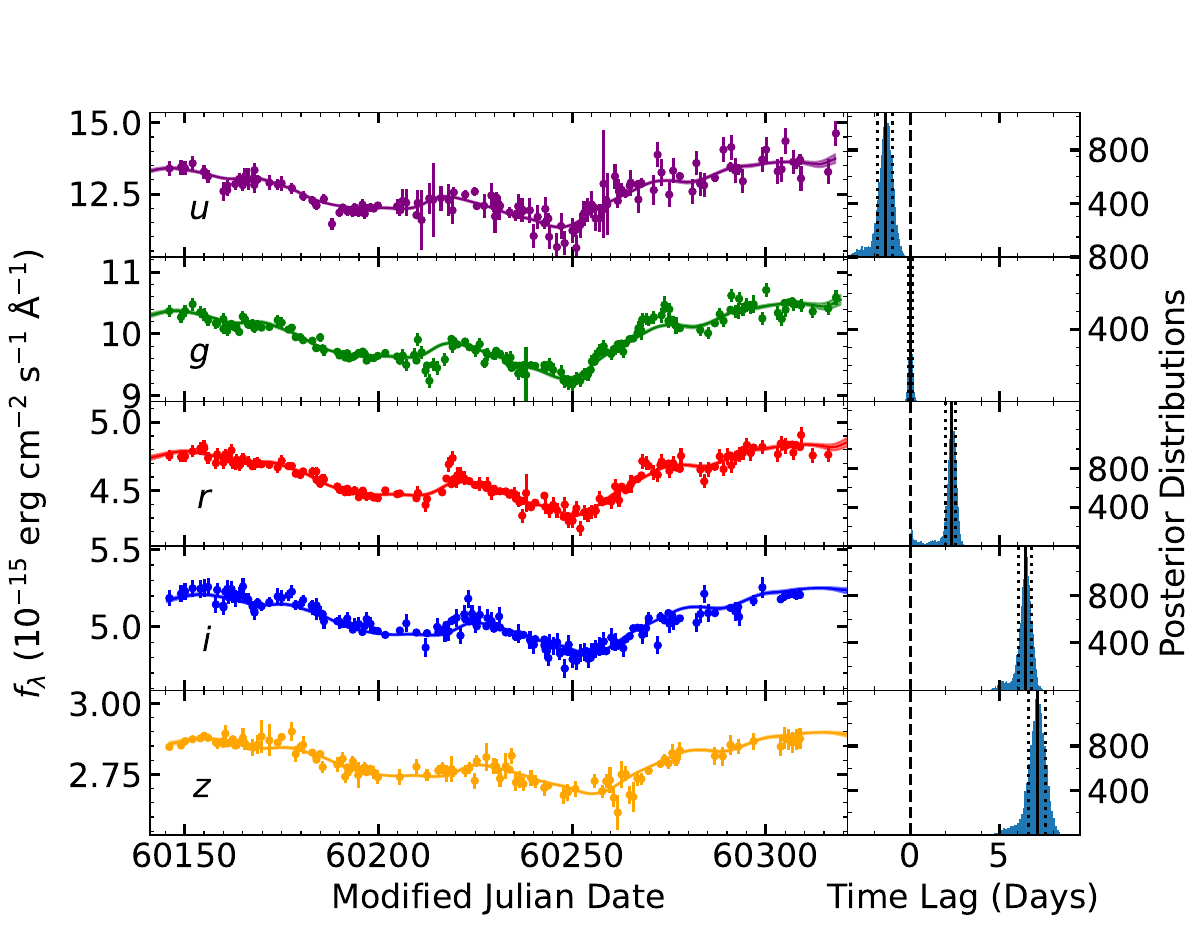}
 \caption{{Light curves of PG~2130+099 with PyROA models overlaid in the $ugriz$ bands. The solid colored lines are the light curve models, with the shaded regions representing the uncertainty. The panels on the right are the posterior distributions of each band, with lags measured with respect to the $g$ band. The solid vertical line is the measured lag, with the dotted lines representing the uncertainty on the lag and the dashed line indicating a lag of 0 days for reference}}
 \label{fig:light curves_PyROA}
\end{figure*}

\begin{figure*}
 \centering
 \includegraphics[width=0.99\textwidth]{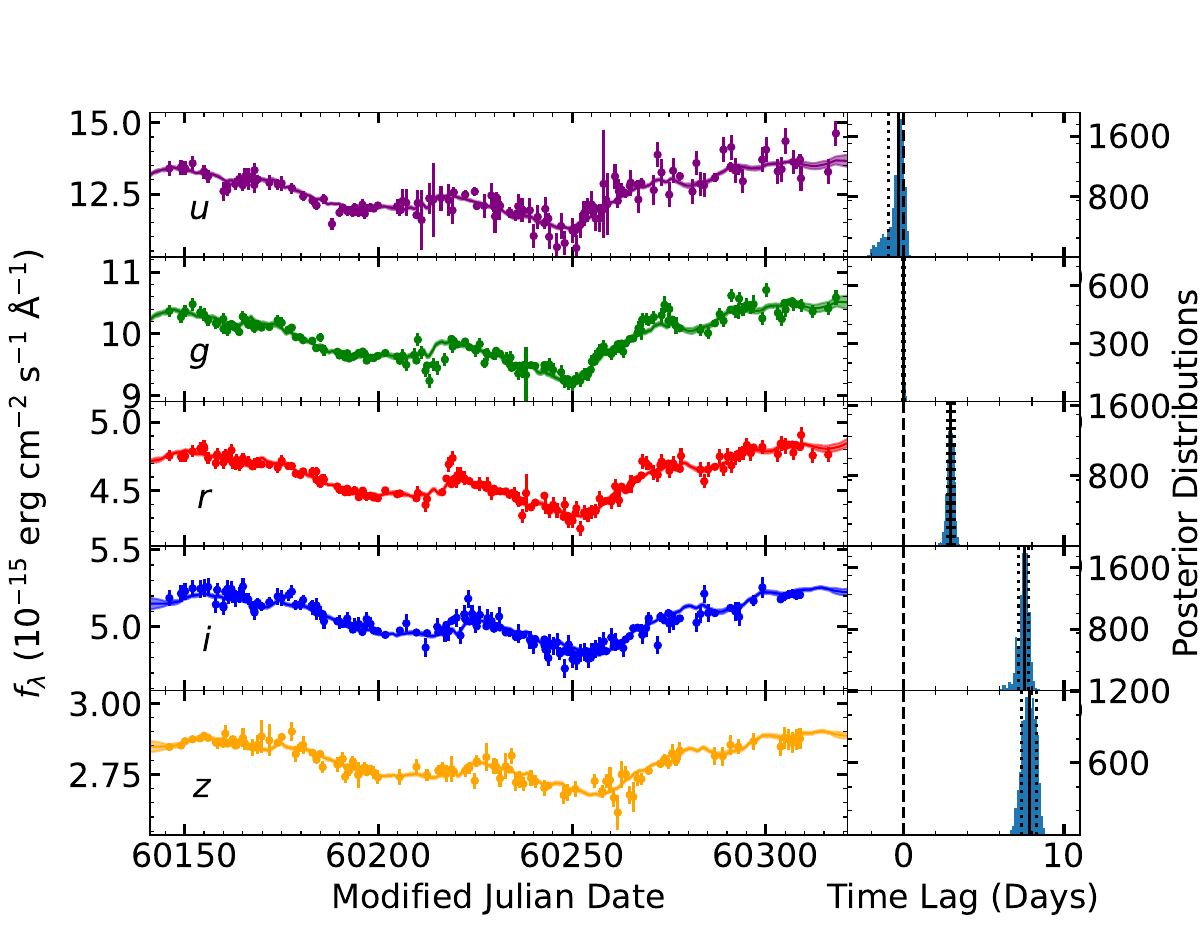}
 \caption{{Light curves of PG~2130+099 with JAVELIN models overlaid in the $ugriz$ bands. The solid colored lines are the light curve models, with the shaded regions representing the uncertainty. The panels on the right are the posterior distributions of each band, with lags measured with respect to the $g$ band. The solid vertical line is the measured lag, with the dotted lines representing the uncertainty on the lag and the dashed line indicating a lag of 0 days for reference}.}
 \label{fig:light curves_JAV}
\end{figure*}

\end{document}

%% file: Lightcurve_table.tex
\begin{deluxetable*}{lcccccc}
\tablecaption{Telescope and Light Curve Statistics
\label{table:telescope_info}}
\tablehead{
Telescope & Total Points & Start Date & End Date & Length (Days) & Average Cadence (Days) & Median Cadence (Days) }
\startdata
ZOWADA & 65 & 60206.28 & 60318.10  & 111.82  & 1.87 & 2.00 \\
LT & 33 & 60146.06 & 60242.85  & 96.79  & 2.51 & 2.00 \\
1m0-01 & 16 & 60224.87 & 60308.82  & 83.95  & 3.80 & 2.00 \\
1m0-03 & 15 & 60160.49 & 60240.48  & 79.99  & 3.86 & 2.00 \\
1m0-08 & 5 & 60158.39 & 60166.39  & 8.00  & 2.00 & 2.00 \\
1m0-10 & 5 & 60179.85 & 60269.76  & 89.91  & 12.24 & 7.00 \\
1m0-11 & 13 & 60167.50 & 60239.39  & 71.90  & 4.00 & 2.00 \\
1m0-12 & 11 & 60168.84 & 60260.76  & 91.92  & 5.60 & 2.00 \\
1m0-13 & 13 & 60169.85 & 60265.78  & 95.93  & 5.00 & 2.00 \\
1m0-14 & 6 & 60267.87 & 60296.85  & 28.98  & 3.90 & 2.00 \\
Combined & 182 & 60146.06 & 60318.10  & 172.04  & 1.48 & 2.00 \\
\enddata
\tablecomments{Light curve Information is for the $g$-band lightcurve. Dates are listed as MJD.}
\centering
\end{deluxetable*}

%% file: Lagtable.tex
\begin{deluxetable*}{lccccccc}
\tablecaption{Time Lags
\label{table:lags}}
\tablehead{
Method & $u$ & $g$ & $r$ & $i$ & $z$ & $\tau_0$ & $\tau_0$ (No $i$ band) }
\startdata
PyCCF
 & $-1.36^{+1.26}_{-1.13}$
 & $0.00^{+0.55}_{-0.56}$
 & $2.34^{+0.57}_{-0.59}$
 & $6.44^{+0.88}_{-0.78}$
 & $7.57^{+1.06}_{-1.16}$
 & $6.52 \pm 0.46$ 
 & $5.90 \pm 0.27$
\\
JAVELIN
 & $-0.28^{+0.28}_{-0.62}$
 & $0.00 \pm 0.04$
 & $2.75 \pm 0.20$
 & $7.07^{+0.25}_{-0.33}$
 & $7.37^{+0.41}_{-0.43}$
 & $6.88 \pm 0.72$ 
 & $5.96 \pm 0.61$
\\
PyROA
 & $-1.28^{+0.35}_{-0.41}$
 & $0.00 \pm 0.11$
 & $2.17^{+0.23}_{-0.32}$
 & $6.08^{+0.29}_{-0.37}$
 & $6.69^{+0.42}_{-0.48}$
 & $6.02 \pm 0.50$ 
 & $5.27 \pm 0.23$
\\
\addlinespace
\enddata
\tablecomments{All measurements are given in units of days. Object lags are displayed in rest-frame.}
\centering
\end{deluxetable*}

%% file: FluxFlux_Fnu_PG2130+099__ugriz_Values.tex
\begin{deluxetable*}{lccccc}
\tablecaption{Flux-Flux Analysis Values
\label{table:fluxflux}}
\tablehead{
 & $u$ & $g$ & $r$ & $i$ & $z$ }
\startdata
Max & $4.959 \pm 0.152$ & $7.194 \pm 0.076$ & $5.807 \pm 0.068$ & $8.844 \pm 0.093$ & $6.492 \pm 0.078$  \\
Mean & $4.238 \pm 0.012$ & $6.644 \pm 0.004$ & $5.429 \pm 0.004$ & $8.46 \pm 0.006$ & $6.256 \pm 0.006$  \\
Min & $3.596 \pm 0.15$ & $6.167 \pm 0.068$ & $4.996 \pm 0.062$ & $7.95 \pm 0.101$ & $5.854 \pm 0.141$  \\
Constant & $0.245\pm0.223$ & $2.374\pm0.097$ & $2.216\pm0.085$ & $5.065\pm0.132$ & $3.87\pm0.105$ \\
RMS & $0.227 \pm 0.013$ & $0.243 \pm 0.006$ & $0.182 \pm 0.005$ & $0.193 \pm 0.008$ & $0.137 \pm 0.006$ \\
\enddata
\tablecomments{Values found from the flux-flux analysis. All values are reported in units of mJy.}
\centering
\end{deluxetable*}

%% file: DiskSizeTable_Update.tex
\begin{deluxetable*}{ccccc}
\tablecaption{PG2130+099 Luminosities and Implied Disk Sizes
\label{table:disksize_lum}}
\tablehead{
Study & Host-Subtracted 5100\AA~Luminosity & 5100\AA~Implied Disk Size (Days)}
\startdata
Jha et al. (2022) & 2.27$\times 10^{44}$ $\pm$ 3.97$\times 10^{43}$ & 2.06 $\pm$ 0.63 \\
Fian et al. (2022) & 2.27$\times 10^{44}$ $\pm$ 3.97$\times 10^{43}$ & 2.80 $\pm$ 0.94 \\
Miller et al. (in prep) & 2.70$\times 10^{44}$ $\pm$ 1.37$\times 10^{43}$ & 6.36 $\pm$ 1.26 \\
Miller et al. (in prep) & 2.52$\times 10^{44}$ $\pm$ 1.88$\times 10^{43}$ & 3.60 $\pm$ 2.58 \\
This Work & 3.12$\times 10^{44}$ $\pm$ 1.87$\times 10^{43}$ & 7.74 $\pm$ 0.54 \\
This Work (i band excluded) & 3.12$\times 10^{44}$ $\pm$ 1.87$\times 10^{43}$ & 7.00 $\pm$ 0.32 \\
\enddata
\tablecomments{Luminosity and 5100$\Angstrom$ Continuum Emitting sizes as presented in Fig.\ref{fig:previous_disk_size_comp}. All luminosities have been adjusted for redshift z~$=0.063$, dust extinction assuming an E(B-V) of 0.037, and adjusted to all use a luminosity distance of 274.7~Mpc. \cite{fian2022} reports a host-subtracted restframe luminosity and performs their own dust correction, so we do not apply these corrections to their value. They use a slightly different measurement for luminosity distance, so we recalculate their luminosity using our value for luminosity distance instead. We use Fian et al.'s luminosity measurement for \cite{jha2022}, as Jha et al. does not calculate the 5100$\Angstrom$ luminosity and both observation campaigns observed PG2130+099 at the same time.  
A host galaxy background luminosity of $3.30\times10^{43}\pm1.65\times10^{42}$~erg/s has been subtracted for both years from Miller et al. (in prep) and this work. All implied continuum emitting sizes are calculated from lags adjusted to restframe.}
\centering
\end{deluxetable*}